\documentstyle[prl,aps,psfig,floats,twocolumn]{revtex} 
\begin{document}
\draft
\twocolumn[\hsize\textwidth\columnwidth\hsize\csname
@twocolumnfalse\endcsname
\title{Theory of Boundary Effects in Invasion Percolation}
\author{Andrea Gabrielli}
\address{Dipartimento di Fisica Universit\`a degli Studi ``Tor Vergata'',
v.le della Ricerca Scientifica 1, 00133 Roma Italy}
\address{Dipartimento di Fisica e INFM - Unit\`a di Roma 1, 
Universit\`a degli Studi ``La Sapienza'' 
P.le A. Moro 2, 00185 Roma, Italy}
\author{Raffaele Cafiero}
\address{Max-Planck Institute for Physics of Complex Systems,
N\"othnitzer Strasse 40, D-01187 Dresden, Germany}
\address{Dipartimento di Fisica e INFM - Unit\`a di Roma 1,
 Universit\`a degli Studi ``La Sapienza'' 
P.le A. Moro 2, I-00185 Roma, Italy}
\author{Guido Caldarelli}
\address{Theory of Condensed Matter Group, Cavendish Laboratory, 
Madingley Road CB3 0HE Cambridge, UK}
\address{Department of Theoretical Physics, The University, M13 9PL 
Manchester, UK}
\maketitle
\begin{abstract}
We study the boundary effects in invasion percolation with and without 
trapping. We find that the presence of boundaries introduces a new set of 
surface critical exponents, 
as in the case of standard percolation. Numerical simulations show
a fractal dimension, for the region of the percolating cluster 
near the boundary, remarkably different from the bulk one. 
In fact, on the surface we find a value of $D^{sur}=1.65 \pm 0.02$ 
(for IP with trapping $D_{tr}^{sur}=1.59 \pm0.03$), compared with 
the bulk value of $D^{bul}=1.88 \pm 0.02$ ($D_{tr}^{bul}
=1.85 \pm 0.02$). We find a logarithmic cross-over from surface to bulk 
fractal properties, as one would expect from the finite-size theory of critical 
systems. 
The distribution of the quenched variables on the growing interface near 
the boundary self--organises into an asymptotic shape characterized by 
a discontinuity at a value
$x_c=0.5$, which coincides with the bulk critical threshold. The exponent
$\tau^{sur}$ of the boundary avalanche distribution for IP without trapping 
is $\tau^{sur}=1.56\pm0.05$; this value is 
very near to the bulk one. Then we conclude   
that only the geometrical properties (fractal dimension) 
of the model are affected 
by the presence of a boundary, while other statistical and 
dynamical properties are unchanged. Furthermore, we are able to present 
a theoretical computation of the relevant
critical exponents near the boundary. This analysis combines two recently
introduced theoretical tools, the Fixed Scale 
Transformation (FST) and the Run
Time Statistics (RTS), which are particularly suited for the study of
irreversible self--organised growth models with quenched disorder. Our
theoretical results are in rather good agreement with numerical data.
\end{abstract}
\pacs{68.70.+W,05.40.+j}
]
\narrowtext
\section{Introduction}

Recently, a large effort has been devoted to the study of Invasion
Percolation (IP)\cite{WW,IPlong,CKLW}.
Compared to standard percolation \cite{stauffer}, IP
has the advantage of describing the dynamical evolution 
of the invading cluster as well as the final result. Furthermore,
since a connectivity condition is naturally implemented in IP, 
its dynamics do not produce extra, undesired, finite clusters, as 
happens instead in standard percolation \cite{stauffer}. 
Even if IP is more difficult to treat theoretically (because  
it presents a non-local, extremal deterministic dynamics in a quenched
disordered medium \cite{IPlong,bak}),
it has been considered the paradigm of a large class of self--organised
critical models .
The Bak and Sneppen model for punctuated equilibrium \cite{bs}, and the 
Sneppen model for surface dynamics \cite{snep} belong to this class. 

In the standard theory of critical phenomena, the role of boundaries 
has been intensively
analysed \cite{DG}, and for many physical situations, ranging from
Ising models to the more recent class of Self Organized Models 
\cite{STC,CTS},
their presence produces a novel set of critical indices related 
to the surface.
The reason for the new behaviour consists in the lack of a 
microscopic layer in the system. This changes dramatically 
the microscopic interactions in the surface region of the system, yielding
eventually to a macroscopically observable characteristic behaviour. 
The standard theory 
of finite size scaling of a thermodynamical system close to its critical
point predicts in two dimensions 
\cite{cardy} a logarithmic cross-over of the critical
exponents from the boundary to the bulk. Consequently, the effect of the
boundary extends over the whole system. This is due to the 
strong correlations peculiar to a critical system.
Such a study has already been done for standard
percolation, and the results are available in the literature 
\cite{cardy1,vander}, but no similar analysis has been performed for IP.
Among the approaches applied to models with extremal dynamics, going 
from Mean Field treatment \cite{bak} to a
recently introduced technique called Run Time Statistics
\cite{IPlong,Matteo}, only the latter one, when combined with the 
Fixed Scale Transformation method \cite{FSTreport}, seems to be 
able to capture the subtle effects due to the presence of a 
boundary in the system.

In this work we present numerical and theoretical evidence that 
a peculiar behaviour on the boundary takes place also in IP. 
Some of the results reported have already been published \cite{ipborap}. 
Here, we would like to give a complete and detailed description 
of numerical results and of the derivation of the analytical results of 
the previous Ref.\cite{ipborap}.
Moreover, we present new analytical and numerical 
results, like the computation of the boundary avalanche
exponent and the  extension of our analysis to the case 
of IP with trapping, which has no analogue in the standard 
percolation model. In particular, the results for IP with trapping 
have no counterpart in standard percolation theory \cite{stauffer}.

From a qualitative point of view, the analogy between boundary effects 
in ordinary critical phenomena and IP can be easily understood by 
considering that boundary sites have 
fewer neighbors than bulk ones and hence fewer chances to invade a new region.  
Moreover, IP is a self--organised critical model and, as the evolution
time tends to infinity, it can be considered in the same way as 
an ordinary thermodynamical system when the temperature is tuned at the
critical value $T_c$.
The crossover between boundary and bulk fractal properties is shown 
by considering intersections of the percolation cluster 
with straight lines parallel to the external boundary.
This subset of the percolating cluster has a fractal dimension
 that varies with the distance from the boundary.
Using some theoretical tools introduced for the study of fractal growth 
processes, the Run Time Statistics (RTS) \cite{IPlong,Matteo} and 
the Fixed Scale Transformation (FST) \cite{FSTreport}, we are able to 
study analytically this behaviour, with an estimation of 
the boundary fractal dimension that is in rather
good agreement with the numerical value. This is done for IP with and
without trapping. In addition we study the avalanche 
dynamics near the boundary, for IP without trapping, and we compute 
both numerically and analytically, by using the RTS and FST schemes, 
the boundary avalanche exponent $\tau^{sur}$.

Our results are presented in the following order.
In section II, we present the definition of the model and 
a review of the numerical data. 
In section III, we describe the concepts underlying
 RTS and FST. In section IV we apply these methods to the computation of the
boundary fractal dimension. In section V we compute the boundary 
avalanche exponent.
In the last part we give a summary of the main topics.Appendix A is devoted
to the derivation of the RTS equations.

\section{The Invasion Percolation Model}
IP was introduced more than 10 years ago \cite{WW} in order to 
describe the slow capillary displacement of a fluid (e.g. oil),
the {\em defender}, 
from a random porous medium due to 
another immiscible invading fluid (e.g. water), the {\em invader}.
 
In general two cases are studied: 1) 
the medium is filled with an 
incompressible defender (e.g. oil), 
which is immiscible with the invader fluid; 
2) the medium is filled 
with a defender with an infinite compressibility. 
In the former case the invader may {\em trap} regions 
of the defender: e.g. as the water advances, it can completely surround 
regions of the oil. These regions become disconnected from 
the other bonds occupied by the defender and, due to 
incompressibility, they become forbidden to the invader. 
This {\em trapping} effect lowers the fractal 
dimension of the percolating invader cluster. 
 From an experimental point of view, 
{\em trapping} is connected to the phenomenon of ``residual oil'', 
which is a great economic problem in the oil industry \cite{sahimi}.
 
The random medium is represented by a network of bonds corresponding to
the throats connecting the pores of the medium.
Let us assume, now, that the invader begins to displace the defender.
Under the condition of a low and constant flow rate, the interface can be 
considered to move one step at 
time, by invading the throat with the smallest section, 
i.e. the throat where there is the largest capillary force \cite{WW}. 
One can mimic this behaviour by 
assigning a random section $x_i$ (here we take an uniform distribution 
in $\left[0,1\right]$) to each
bond $i$ of the medium. The invading cluster evolves by occupying the bond
with the smallest $x_i$ on its perimeter. This is what is called 
a {\em deterministic extremal dynamics}.

\begin{figure}
\centerline{\psfig{figure=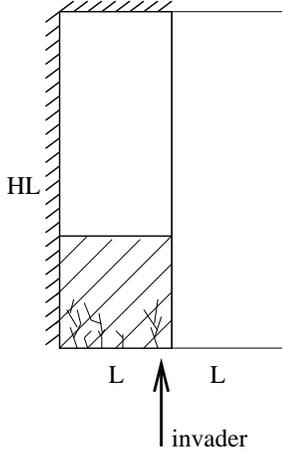,height=6cm}}
\caption{Set up of numerical simulations. An invading (not 
yet percolating), cluster is shown. Only the bottom left part 
of the cluster will be considered for the statistics.}
\label{fig1}
\end{figure}

To study the behaviour at the boundary of this model, 
we performed some numerical simulations
in the system shown in FIG.\ref{fig1}, representing a sample of a 
two dimensional square lattice. 
To study the effect of only one boundary (e.g. the left one), 
we ensured isolation from the other one. 
To obtain this, we choose a lattice with size $HL \times 2L$ where
$H=2,3,4$, and the initial invader cluster is composed of the first L 
bonds of the bottom line, starting from the left boundary. 
The simulation stops when the cluster percolates the system, i.e. 
when the growth reaches the top of the sample.

\begin{figure}
\centerline{\psfig{figure=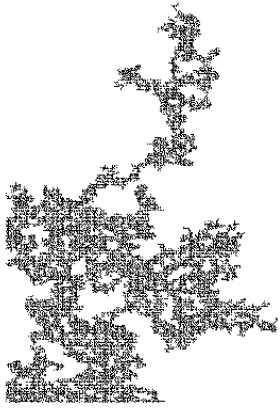,height=6cm}}
\caption{This picture shows the entire cluster. 
The region of interest in
which statistics is taken is the lower-left one.}
\label{fig2}
\end{figure}

\begin{figure}
\centerline{\psfig{figure=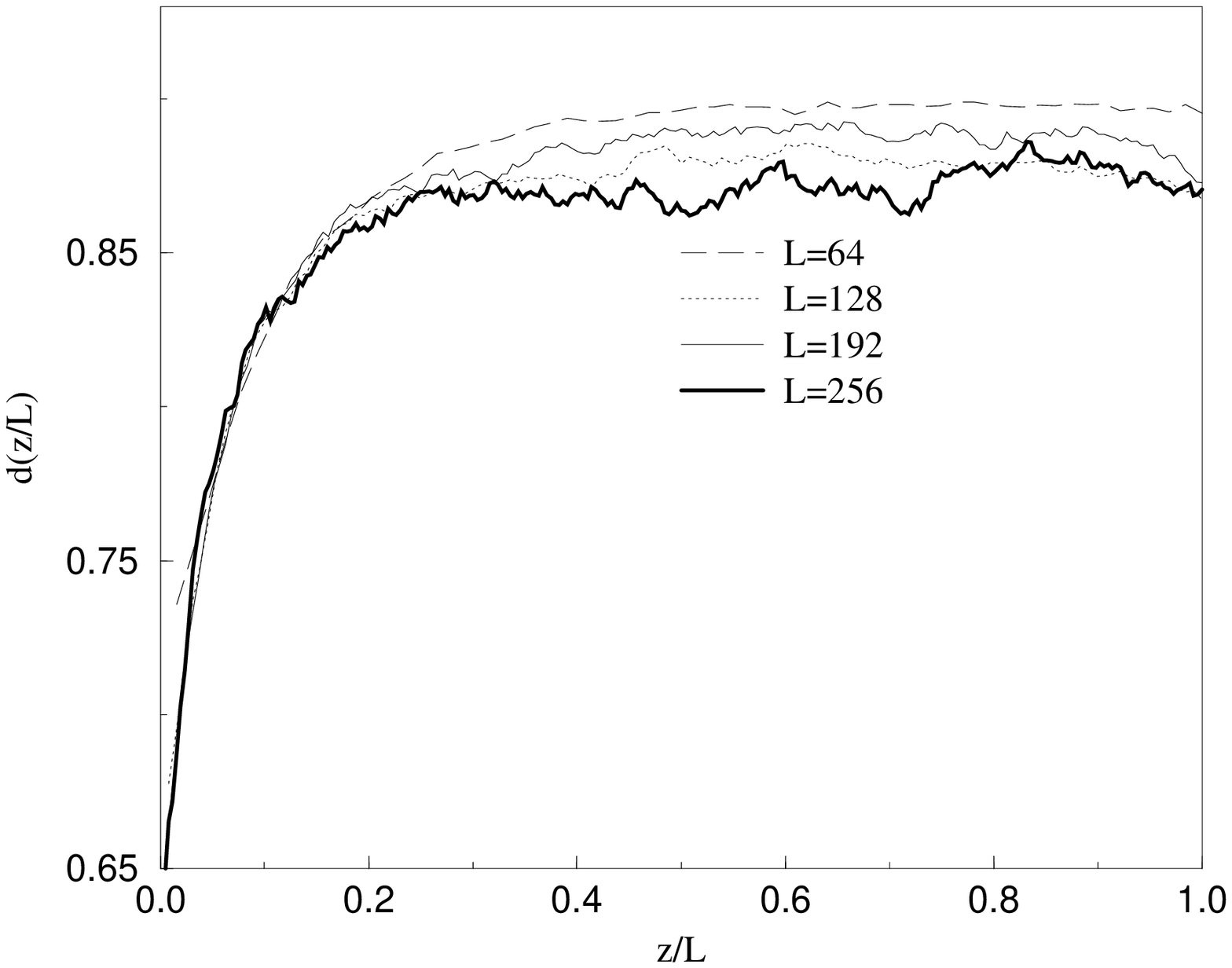,height=6cm}}
\caption{Behavior of the fractal dimension 
of the intersection set of IP without trapping 
versus the normalized distance $z/L$ from 
the boundary ($H=4$). $z$ and $L$ are in 
lattice units, then the normalized 
distance is dimensionless.}
\label{fig3}
\end{figure}

In FIG.\ref{fig2} a typical realization of this process is shown. 
The region of interest is the bottom-left one in FIG.\ref{fig1}, where 
we can assume that the region is ``frozen" with respect to the 
invasion process, i.e. the asymptotic fractal properties of the percolating 
cluster are well defined. 
For each value of the system size $L$ we collected a set of $10^3$ 
different realizations. 
In the region  where statistics is collected, we study the fractal 
dimension of the sets of points obtained by intersecting the percolating
cluster with lines parallel to the boundary, at a distance $z$ from it.
In this way, we are able to follow the cross-over 
of the fractal dimension of 
the cluster from the boundary to the bulk region. 
A standard box-counting procedure 
is used to compute the fractal dimension of the intersections. 
The behaviour of the
fractal dimension $d(z/L)$ of the intersections as a function of the
normalized distance $z/L$ from the left boundary is 
presented in FIG.\ref{fig3} for $L=256, H=4$ (i.e. $512 \times 1024$).
In Table \ref{tab3} we present the values of the boundary fractal dimension
$d^{sur}$ for different system sizes $L$ and different values of $H$.
For the largest simulation $L=256$, $H=4$ (i.e. $512 \times 1024$), 
we obtain the result that the fractal dimension of this subset of the 
cluster passes from $d^{sur}=0.65 \pm 0.02$ on the boundary 
to $d^{bul}=0.88 
\pm 0.02$ in the bulk (at a distance $z/L \sim 0.4$ from the boundary), 
where $d^{bul}$ represents the fractal dimension 
of the intersection far away from the boundary. 
A similar behaviour holds for smaller sizes $L$ and $H$ as well.
Since the dimension of the intersection set $d$  obeys 
$d=D-1$ where $D$ is the fractal dimension of the cluster, 
the last result is in agreement with the known value of $D\simeq 1.89$. 

\begin{figure}
\centerline{\psfig{figure=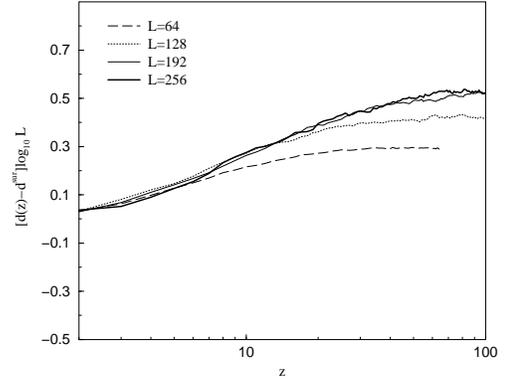,height=6cm}}
\caption{
Collapse plot of $[d(z)-d^{sur}]log(L)$ for the 
different sizes, in $log-linear$ scale, 
for IP without trapping ($H=4$). 
$z$ and $L$ are in lattice units, 
then the normalized distance is dimensionless.}
\label{fig4}
\end{figure}

\begin{figure}
\centerline{\psfig{figure=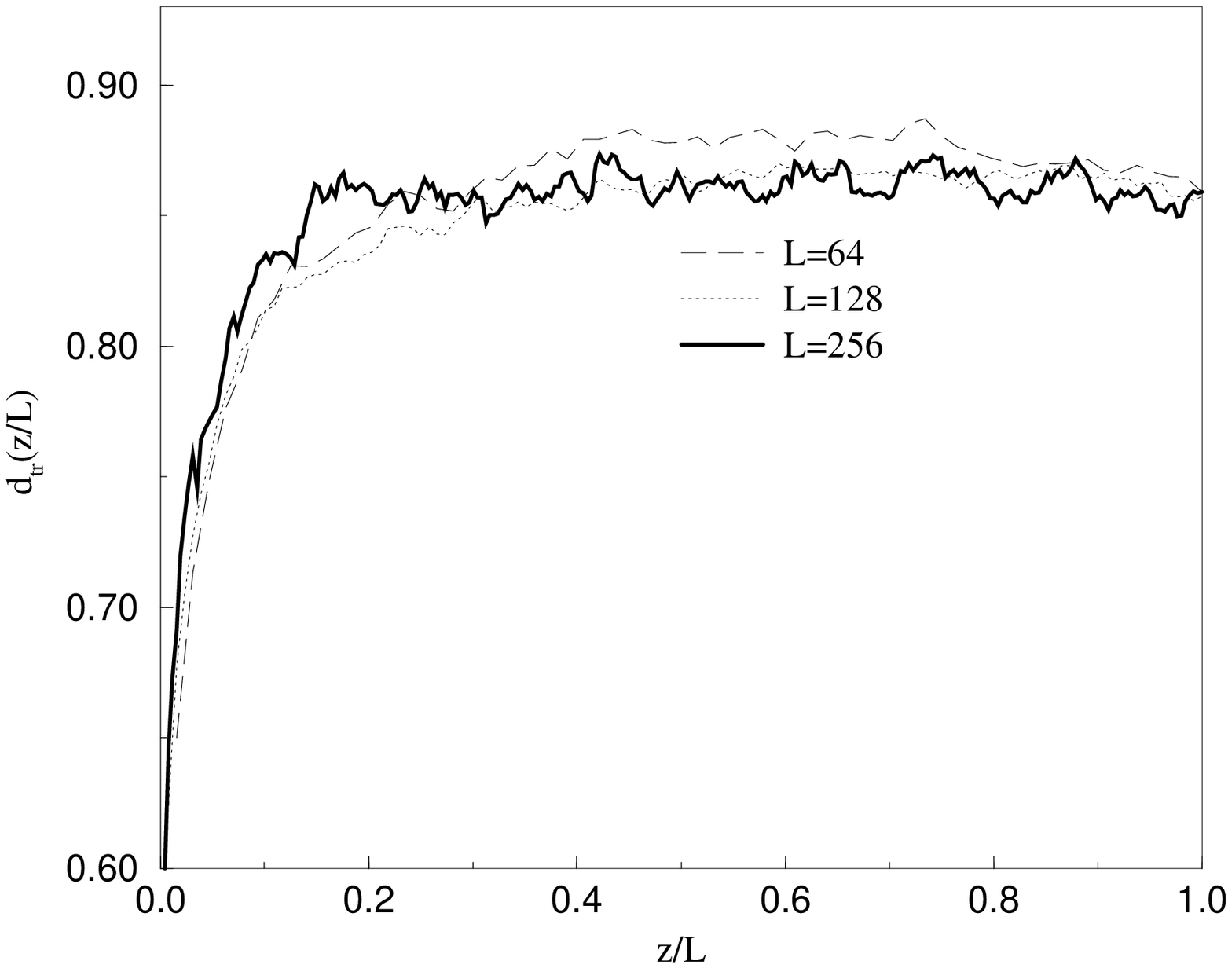,height=6cm}}
\caption{Behavior of the fractal dimension 
of the intersection set of IP with trapping
versus the normalized distance $z/L$ from 
the boundary ($H=4$).
$z$ and $L$ are in lattice units, 
then the normalized distance is dimensionless.}
\label{fig4a}
\end{figure}

In order to explain such a slow crossover 
from $d^{sur}$ to $d^{bul}$ we assumed that the number of
occupied sites $N(z,L)$ at a distance $z$ from the boundary 
follows the finite-size scaling law:
\begin{equation}
N(z,L)=L^{d^{bul}} f(z/L)
\end{equation} 
where one has 
$f(z/L)\sim (z/L)^{(d^{bul}-d^{sur})}$ for $z<<L$ 
and $f(z/L)=\em{const}$ for 
$z>>L$. 
Then in the first region we should have:
\begin{equation}
d(z)=d^{sur}+(d^{bul}-d^{sur})log(z)/log(L)
\end{equation} 
To test this scaling hypothesis 
we collapsed the curves relative to different $L$ by plotting
$\left[d(z)-d^{sur}\right]log(L)$ as a function of $z$. 
The result depicted in FIG.\ref{fig4} shows a rather good collapse 
in the small $z$ region.
A similar behaviour is found for IP with site trapping.
In order to implement site trapping in our simulations, 
after each growth step a fictitious Laplacian field $\phi$ is relaxed on the 
growing structure, with the following boundary conditions: $\phi=0$ on
the bottom boundary, the left boundary and the invading cluster, while 
$\phi=1$ on the top boundary.
In this way, all the bonds in a closed, trapped 
region are characterized by $\phi =0$. Then it is possible to 
recognize trapped bonds and to eliminate them from the list of
bonds allowed to grow at the next step. 
Obviously, in this case the numerical simulations need
much more time to be performed and we have been 
able to collect a smaller, but
still significative, statistics with respect to IP without trapping ($10^3$
clusters for $L=64$, $2\times10^2$ clusters for $L=128$ and $10^2$ clusters 
for $L=256$ each one for $H=2,3,4$). 
In FIG. \ref{fig4a} we show the behaviour of the intersection 
dimension $d_{tr}(z/L)$ versus the normalized distance from the boundary, 
each simulation is for a values of $H=4$. The
fractal dimension $d_{tr}$ is computed on samples $512 \times 1024$ 
(i.e. $L=256$ and $H=4$) and passes from $d_{tr}^{sur}=0.59\pm0.03$ 
on the boundary to $d_{tr}^{bul}=D_{tr}^{bul}-1=0.85\pm0.03$ in the bulk,
which is agreement with the known value $D_f\sim1.82$ for site 
trapping \cite{WW}. 
The data shown in Table \ref{tab4} exhibit the same slow logarithmic 
cross-over found for IP without trapping.

Other important quantities characterizing the dynamical properties of 
IP are the average distribution of quenched variables 
on the perimeter, called {\em histogram} $\Phi_t(x)$, which gives an evidence
of the self--organised nature of the model, and the avalanche-size
distribution in the asymptotic critical state $Q(s;x_c)$, where $x_c$ is 
the "self-critical" threshold of the model.

\begin{figure}
\centerline{\psfig{figure=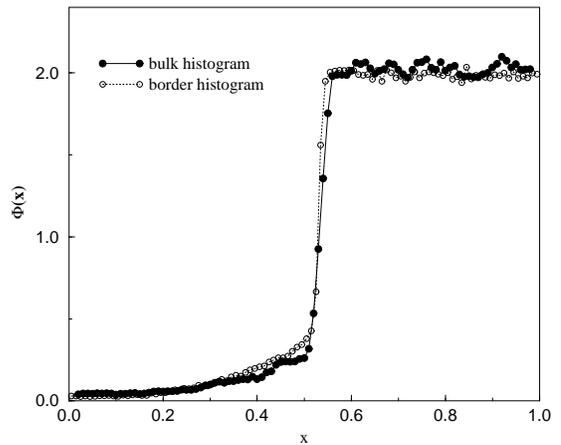,height=7cm}}
\caption{The histogram distribution of bulk 
perimeter variables in IP without trapping is compared
with the distribution of variables near the 
boundary of the system, after $5 \times 10^3$ time steps. The two 
distributions coincide.}
\label{fig5}
\end{figure}

Let us start with the study of the histogram for IP without trapping.
It is known\cite{Matteo} that for the bulk IP, the {\em histogram} 
distribution evolves in time from the initial flat shape, and 
self--organises into a step function with a discontinuity at 
a critical value $x_c^{bul}$ which depends on the details of the model and 
on the embedding dimension \cite{WW} and coincides with the critical threshold
of the classical percolation in the same kind of lattice.
For the two dimensional square lattice one has 
$x_c^{bul}=\frac{1}{2}$. 
Our simulations show that the distribution 
of the $x_i$'s on the boundary self--organises into a theta function
and the critical threshold is again $x_c^{bul}$. 
A comparison between the bulk histogram 
and the boundary one is shown in FIG.\ref{fig5}.
It is not surprising to find a similar behaviour, because the value 
of the boundary critical threshold is dependent on the dynamical evolution 
of the whole percolating cluster. Since for bulk IP the trapping
does not affect the histogram distribution \cite{WW}, 
the introduction of the trapping does not modify the above result.

\begin{figure}
\centerline{\psfig{figure=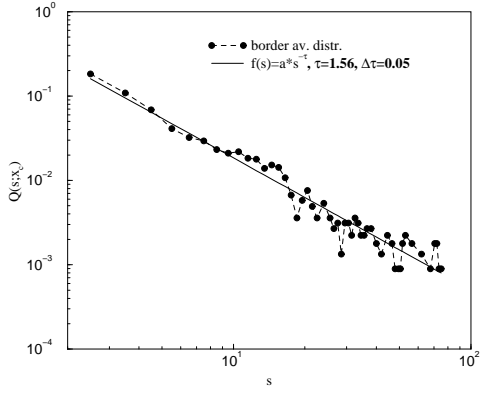,height=6cm}}
\caption{
Border avalanche distribution of IP without trapping 
in $log-log$ scale. 
The least square fit
 gives a slope $\tau^{sur}=1.56\pm0.05$.}
\label{fig6}
\end{figure}

Another important quantity describing the dynamics of the model is 
the critical avalanche-size distribution $Q(s;x_c)$.
An avalanche is a sequence of elementary growths events 
causally and geometrically connected to a first one, which is called 
the {\em initiator} of the avalanche.
That is, if one consider an event of growth of 
the {\em initiator} (a certain bond $k$),
the avalanche lasts until the bonds selected to grow are those joining  
the growth interface after the growth of bond $k$ (bonds ``younger" 
than $k$).
Note that all these bonds have the related random number $x$ smaller than 
the {\em initiator} one.  
If the bond selected by the dynamics was on the perimeter before
the initiator growth, then the avalanche stops.
In the asymptotic limit, due to the step shape of the histogram, only bonds
with $x\le x_c$ grow.
We call $Q(s;x)$ the size distribution of avalanches whose initiator is 
associated with a number equal to $x$.
It is shown for bulk IP, both through numerical simulations and 
theoretical calculation,
$Q(s;x)$ is scale invariant (i.e. is a power law), only  
if the variable $x$ of the {\em initiator} is equal to $x_c$. 
If $x<x_c$ $Q(s;x)$ has an exponential cut off at a typical size 
$s_0\sim (x_c-x)^{-\sigma}$ with $\sigma>0$.
The avalanche distribution $Q(s;x_c)$ for bulk IP without trapping has a
power law shape with an exponent $\tau^{bul}\sim 1.60\pm 0.02$ \cite{IPlong,maslov}.
The dynamical activity near the boundary can be characterized by 
the distribution of the avalanches whose first bond (initiator) is located 
on the boundary. 
We performed a set of about $10^3$ numerical simulations of IP without 
trapping, 
of size $3L\times5L$ with $L=128$, lasting $4 \times 10^5$ time steps and 
collected 
the statistics of boundary avalanches from the last $2 \times 10^5$ time steps, in order to ensure
that the system is in its asymptotic critical state. To
identify the single avalanche, we followed \cite{IPlong}, by adding the
condition that the initiator of the avalanche is on the boundary. 
In FIG.\ref{fig6}
we show the behaviour of the boundary avalanche distribution. We find:
$\tau^{sur}=1.56\pm0.05$. This value is very near to the bulk value, and we
can conclude from our numerical analysis that bulk and boundary avalanches 
have the same distribution. In section V we will derive 
analytically this result.

\section{Run Time Statistics and Fixed Scale Transformation}

In this section we introduce the theoretical tools we used to compute 
the boundary fractal dimension $d^{sur}$ of the infinite IP cluster and 
the boundary avalanche exponent $\tau^{sur}$. 
Our strategy combines Fixed Scale Transformation (FST) \cite{FSTreport} and 
Run Time Statistics (RTS) \cite{IPlong,Matteo}. We describe
briefly the FST approach and we focus  more on the RTS.

\begin{figure}
\centerline{\psfig{figure=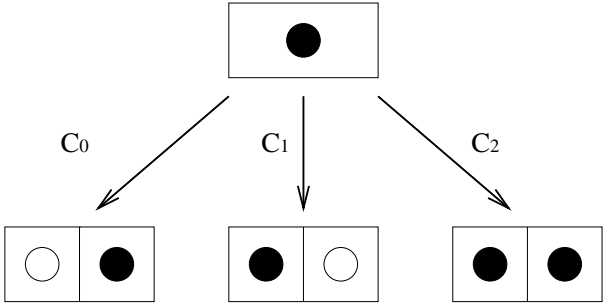,height=4cm}}
\caption{
Fine graining transformation for occupied cells.}
\label{fig7}
\end{figure}

FST is a lattice path integral scheme allowing one to evaluate the 
spatial correlation properties of the intersection between an 
infinite fractal cluster and a straight line. This approach is based on the 
statistical invariance of the correlation properties under a parallel 
translation of the intersecting line (valid for fractals which are 
homogeneous, at least in the translation direction). 
In particular, it is possible to compute the probabilities 
$C_0$,$C_1$ and $C_2$ related to the
configurations $0,1,2$ of the fine graining process of FIG.\ref{fig7}.
For the normalization condition it follows:
\begin{equation}
C_0+C_1+C_2=1
\label{norm}
\end{equation}
 From these probabilities one can compute 
the fractal dimension of the intersection by:
\begin{equation}
d=\frac{log(C_0+C_1+2C_2)}{log 2} = \frac{log(1+C_2)}{log 2}
\label{difra}
\end{equation}
As usual, due to the intersection dimension rule, 
the fractal dimension $D$ of the analysed cluster is given by $D=1+d$. 
The probabilities $C_0, C_1$ and $C_2$ are computed through the statistical 
weights of growth paths, once a
stochastic dynamical formulation of the model is given. This means that 
the use of FST is straightforward whenever a simple calculation 
of the growth paths on the lattice is possible. 
In the present case, there are two problems to overcome in applying the
FST.

Firstly, the fractal properties of the system depend on the distance from 
the boundary ($C_0=C_0(z), C_1=C_1(z),C_2=C_2(z)$), 
this extrapolation from the intersection dimension to the global dimension 
is no more allowed. Moreover, what we actually can compute 
with the FST method are the local (near to the boundary) correlations
orthogonal to the boundary, while the fractal dimension 
of the intersection set parallel to the boundary is given by the 
correlation properties parallel to the boundary. 
However, since the crossover of the fractal dimension 
from the boundary to the bulk is very slow (logarithmic), one is allowed  
to assume that the cluster is "locally" isotropic. 
In this case transversal and horizontal correlations in a thin 
(with respect to the system size) strip parallel to the boundary share
similar properties. 
For the same reason, we can evaluate the fractal dimension $d$ of the 
intersection between the cluster and a straight line parallel to the lateral
boundary, through the first neighbors correlations orthogonal to the same 
boundary at the same distance.

Secondly for IP (and for any other model with deterministic
extremal dynamics) the calculation of the growth paths is extremely difficult, 
because the weight of a path cannot be written as 
the product of the probabilities of the single steps composing it. 
The extremal dynamics of IP is {\em deterministic}, and the disorder
appears only as a realization of quenched random variables.
This implies that to evaluate the statistical weight of a given path  
we have to perform an average over all the quenched disorder and 
this average does not factorize itself in the product of the averages of the 
single steps composing the path. 
The latter problem is solved by the introduction of the RTS transformation.  
This transformation allows us to represent a quenched-extremal process
like a stochastic dynamics. 

As regards the RTS (for a more detailed discussion see \cite{IPlong}), the
starting point is, at each time step $t$, to consider an effective 
probability density $\rho_{i,t}(x)$ for the random number $x_i$ 
associated to each bond $i$ of the growing interface $\partial{\cal C}_t$. 
This density depends on the growth history of the dynamics. 
In fact, $\rho_{i,t}(x)dx$ gives the probability that the variable $x_i$
for the bond $i$ at time $t$ is in the interval $\left[x,x+dx\right]$, 
conditioned by the past growth dynamics of the cluster. 
If a bond $i$ does not belong to the cluster, or to the growth interface,
its effective probability density is the flat one.
Meanwhile, the bonds on the growth interface show a more interesting 
form of distribution.
Once the densities $\rho_{i,t}(x)$ for each bond $i$ 
on the interface are known, one 
can calculate the growth probability distribution $\{\mu_{i,t}\}$ 
(i.e. the probability of being the minimum on the interface) 
at that time step 
for each interface bond (see appendix A):

\begin{equation}
\mu_{i,t}=\int_0^1 dx \rho_{i,t}(x) \prod_{j \in \partial{\cal C}_t-\{i\}}
\left[ \int_x^1 dy \rho_{j,t}(y) \right] .
\label{mu}
\end{equation}
where $\partial{\cal C}_t-\{i\}$ represents the growth interface 
at time $t$ except for the bond $i$.
The effective probability density of any surviving 
bond $j$ at time $t+1$ on the interface must then be updated, conditioned
to the previous growth history at time $t$, i.e. the growth 
of the bond $i$. 
The corresponding equation is (see appendix A):
\begin{equation}
\rho_{j,t+1}(x)=\frac{\rho_{j,t}(x)} {\mu_{i,t}} \int_0^x dy \rho_{i,t}(y) 
\prod_{k\in \partial{\cal C}_t-\{i,j\}}
\left[ \int_y^1 dz \rho_{k,t}(z) \right] 
\label{mu2}
\end{equation}
where $\partial{\cal C}_t-\{i,j\}$ is the growth interface except 
for bonds $i$ and $j$.
New bonds added to the 
perimeter are assigned an effective probability density 
according to a uniform distribution in $\left[0,1\right]$,
as no information is available about them till this time step. 

The above formalism allows us to write the statistical weight of a path 
as the product of the probabilities of individual steps. 

\section{Computation of the boundary fractal dimension}

\begin{figure}
\centerline{\psfig{figure=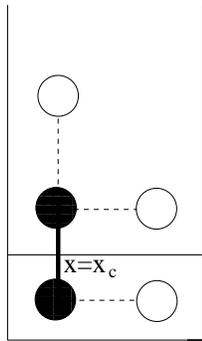,height=4.5cm,angle=-90}}
\caption{
The growth column used in the FST scheme. The thick 
bond is the initiator
 of the avalanche, with $x=x_c$, and the dashed 
bonds are the bonds of the perimeter after the initiator's growth.}
\label{fig8}
\end{figure}

In order to combine the FST and the RTS approach, 
we need to have scale invariant growth rules
(we want to compute a critical exponent, the fractal dimension, 
and the result
cannot depend on the scale). The extremal dynamics of IP is known to be 
independent on the choice of the initial distribution of quenched
variables. 
By using this symmetry, one can show that the scale invariant dynamics 
for block variables is identical to the microscopic dynamics \cite{IPlong}. 
The FST performs the computation of the
correlation properties of a given structure by considering only the growth 
processes inside a growth column (FIG. \ref{fig8}). 
This approximation has been shown to be a good one for the dielectric 
breakdown model (DBM)
\cite{NPW}, and for bulk IP \cite{IPlong}. 
Since eqs. (\ref{mu}), (\ref{mu2}) involve all the variables on the
perimeter of the growing cluster,
a limitation of the process in the growth column destroys these correlations, 
leading to compact clusters \cite{IPlong}. 
The solution to this problem is given by observing that, as the critical 
avalanche size distribution is a power law,
the statistical properties of a generic one (i.e. an 
avalanche whose initiator $i$ has $x_i=x_c$) are then scale invariant. 
Then if one considers the dynamical evolution of a generic critical avalanche 
inside the growth column one obtains the scale invariant correlation 
properties (i.e. $C_0$, $C_1$ and $C_2$) needed to compute 
the fractal dimension.   
This can be done by modifying
the eqs. (\ref{mu}), (\ref{mu2}), in order to take account the 
dynamical evolution of a single critical avalanche. 
We consider a growth column on the perimeter of the infinite structure 
($t\rightarrow\infty$). The starting point is 
the observation that scale invariant asymptotic avalanches begin with an
initiator at $x=x_c$, due to the asymptotic shape of the histogram. 
All the memory of the past growth history is then contained
in the requirement that the initiator has $x=x_c$. Then one is allowed 
to consider explicitly only the bonds grown after the growth of the initiator.
 
The RTS dynamics corresponding to the 
local scale invariant dynamics, is obtained by 
\begin{itemize}
\item{considering only bonds inside the growth column;}
\item{imposing that any ``active" bond $i$ in the column can grow 
only if the value of its variable $x_i$ is less than $x_c=1/2$. 
The idea is that if $x_i>x_c$ for all the 
bonds in the growth column, growth will occur 
at some other place in the structure outside the growth column 
(it coincides with the definition of scale invariant avalanche);}
\item{imposing that the initial bond (i.e. the {\em initiator}), 
which is the largest of the variables participating to the growth process,
 has exactly $x_i=x_c$}.
\end{itemize}
In this way we modified the Eqs. 
(\ref{mu}) and (\ref{mu2}),  
limiting the product over the perimeter variables to variables inside
the growth column.

Because of the presence of a lateral surface, this model is 
intrinsically anisotropic, and consequently we have to introduce some 
modification to the usual way of performing the FST for the bulk IP. 
The anisotropy of the environment implies a breaking of symmetry in the 
FST basic configurations in FIG.\ref{fig7}. Then, due to the presence of the 
boundary, the probabilities $C_0$ and $C_1$ are not equal in this case.

Through the FST one may compute directly the matrix elements $M_{ij}$ and 
from the relation:
\begin{equation}
\left(
\begin{array}{c}
C_0 \\
C_1 \\
C_2 \\
\end{array}
\right)
=
\left(
\begin{array}{ccc}
M_{00}  &  M_{10} & M_{20} \\
M_{01}  &  M_{11} & M_{21} \\
M_{02}  &  M_{12} & M_{22} \\
\end{array}
\right)
\left(
\begin{array}{c}
C_0 \\
C_1 \\
C_2 \\
\end{array}
\right)
\label{FSTmatr}
\end{equation}
it is possible to evaluate $C_2$ and, by using Eq.(\ref{difra}), $d$. 
In this  case 
\begin{equation}
M_{01}=M_{10}=0
\end{equation}
and
\begin{equation}
C_2=\frac{M_{12}M_{02}}{M_{12}+M_{21}(M_{02}-M_{12})
+M_{12}(M_{02}-M_{22})}
\label{FSTsol}
\end{equation}

\begin{figure}
\centerline{\psfig{figure=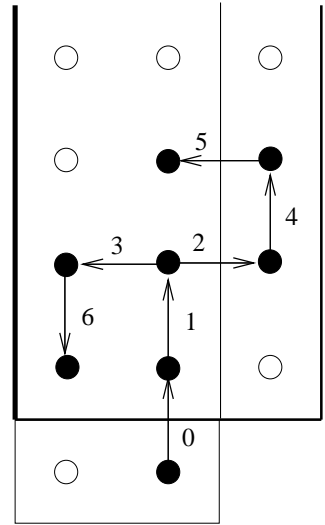,height=7cm}}
\caption{
A possible path of growth at the $6^{th}$ order. 
The invasion proceeds along the arrows, from one black point to 
another one. 
The number near each arrow is the growth time.}
\label{fig9}
\end{figure}

\begin{figure}
\centerline{\psfig{figure=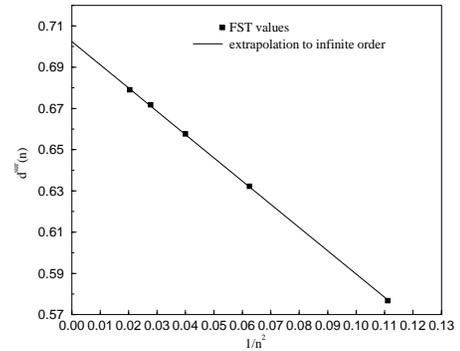,height=5.5cm}}
\caption{
The power law fit we used to get the extrapolated 
value of $d^{sur}$ for IP without trapping.}
\label{fig10}
\end{figure}

The anisotropy of the environment is also introduced in the 
lateral boundary condition of the growth column 
where the FST calculation is performed. 
At the left side of the column we impose the presence of a rigid wall 
and at the right side the paths are allowed to go out and then to return 
inside the growth column, as can be seen in FIG.\ref{fig9}.
In this way we have obtained the results 
shown in Tab.\ref{tab1}, 
where the  fractal dimension for increasing order $n$ (the path length) 
of the FST computation is given. We used a power law fit (FIG.\ref{fig10}) 
to extrapolate $d^{sur}(n)$ to $n=\infty$ and obtained $d^{sur}
(\infty)\simeq 0.70$.

\begin{figure}
\centerline{\psfig{figure=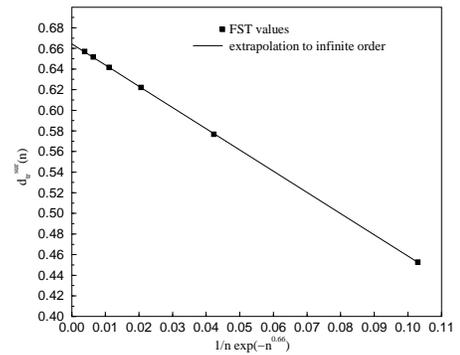,height=5.5cm}}
 \caption{
Extrapolation of the FST fractal dimension 
$d_{tr}^{sur}$for IP with site trapping.}
\label{fig10a}
\end{figure}

A similar approach has been applied to IP with site 
trapping, in particular:
when a growth path produces a closed region surrounding 
the initial pair configuration $C_i$ (see FIG. \ref{fig9}), it stops and
its statistical weight contributes to the matrix elements 
$M_{i,1}$, since the empty right (or left) site above the initial 
configuration cannot be occupied anymore. 
The results are shown in Tab.\ref{tab2} and in FIG.\ref{fig10a}. 
We have extrapolated our results to $n=\infty$ by using the following function 
(see FIG.\ref{fig10a}): 
\begin{equation}
d_{tr}^{sur}(n)=\frac{1}{n} exp(-n^{\alpha})
\end{equation}
with $\alpha=0.66$. This extrapolation gives $d_{tr}^{sur}(\infty)\simeq 0.66$.

\section{Computation of the boundary avalanche exponent}
We now propose a simple theoretical scheme for the analytical 
calculation of the boundary avalanche 
exponent $\tau^{sur}$ of IP without trapping, 
based on the RTS and the FST ideas, which has been 
successfully applied to bulk IP \cite{IPlong}.

The following functional form for the avalanche 
size distribution is assumed:
\begin{equation}
Q(s;x) = s^{-\tau^{sur}}f(|x-x_c^{sur}|s^{\sigma})
\label{disvalanghe}
\end{equation}
where $x_c^{sur}=1/2$ is the critical threshold. 
The function $f(x)$ has the following properties: 
$\lim_{x\rightarrow 0}f(x)=\alpha\neq 0$, and for large values 
of $x$ one has $f(x)\sim e^{-x}$.
Since the size $s$ of the avalanche also includes the initiator, 
the normalization condition for the eq.(\ref{disvalanghe}) is:
\begin{equation}
\sum_{s=1}^{\infty}Q(s;x)=1\;\;\;\forall x\in \left[0,1\right].
\label{normval}
\end{equation}
Usually eq.(\ref{disvalanghe}) holds for $s\gg 1$. 
However, if we consider the dynamics at a certain scale 
$\ell$, we can use eq.(\ref{disvalanghe}) to describe
the statistics of avalanches at that scale.
In the limit $t \rightarrow \infty$, 
for $x=x^{sur}_c$, the asymptotic behaviour described by 
eq.(\ref{disvalanghe}) holds for smaller and smaller values of $s$ as 
$\ell$ is increased. 
The deviations from the pure power law behaviour are integrated out into the
dynamics at scale $\ell$.
For $\ell\gg 1$ we are allowed to suppose that eq.(\ref{disvalanghe}) 
holds from $s=1$ to $s>>1$.
In this case the normalized form of eq.(\ref{disvalanghe}), 
for $x=x^{sur}_c$ is:
\begin{equation}
Q(s;x^{sur}_c)=\frac{s^{-\tau^{sur}}}
{\sum_{s=1}^{\infty}s^{-\tau^{sur}}}.
\label{distrPc}
\end{equation}
The denominator of eq. (\ref{distrPc}) is the {\em Riemann zeta function}, 
$\zeta(\tau^{sur})$.

 From Eq.(\ref{distrPc}), valid if the initiator is at $x_c$ one has:
\begin{equation}
Q(s=1;x^{sur}_c)=\frac{1}{\sum_{s=1}^{\infty}
s^{-\tau^{sur}}}=\frac{1}{\zeta(\tau^{sur})}.
\label{1valPc}
\end{equation}

\begin{figure}
\centerline{\psfig{figure=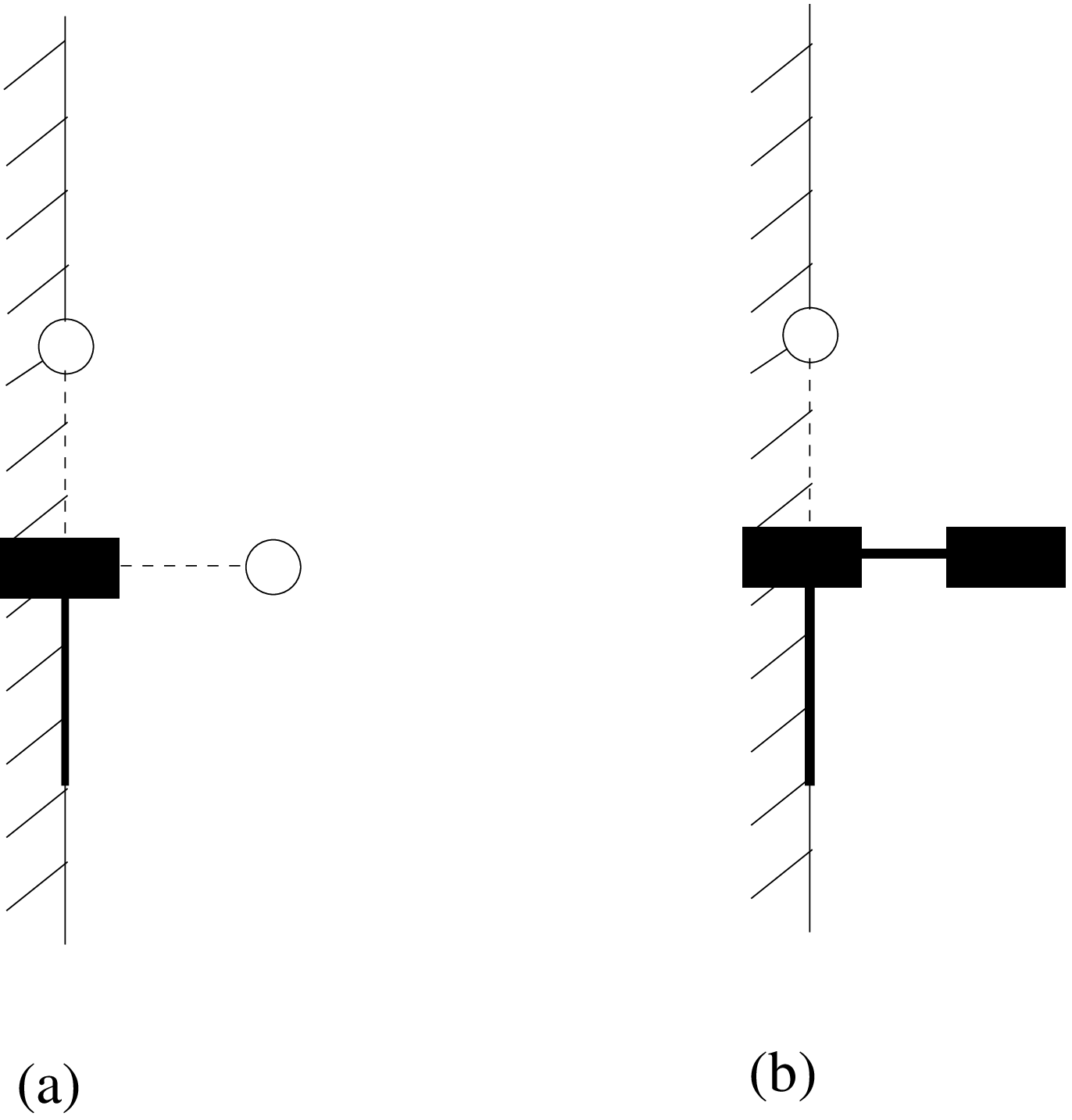,height=7.5cm,angle=-90}}
\caption{(a)-(b): Boundary conditions for a 
 boundary avalanche after the growth of the initiator. 
( ($\bullet$) indicates the cluster sites 
and ($\circ$) the perimeter ones: 
the filled segments are grown bonds and the 
dotted ones are the descendant of
the initiator. The left boundary is shown. Its effect 
is to reduce the maximum number of perimeter bonds 
in which the avalanche can go (2), 
with respect to the bulk (3).}
\label{fig11}
\end{figure}

To obtain an analytic estimation of the boundary avalanche exponent
$\tau^{sur}$ one has to evaluate the left hand side by taking 
into account the boundary conditions near the avalanche, together with 
the presence of the boundary. Then inverting eq.(\ref{1valPc}) it is possible
to measure $\tau^{sur}$.
Let us evaluate $Q(s=1;x^{sur}_c)$. The event $s=1$ means that after the 
growth of the initiator with variable $x^{sur}_c$ the avalanche stops. 
Thus, we consider the initiator $i$ that grew at time $t_0$ and we compute 
the probability that the avalanche stops time $t_0+1$.
This happens when all the descendant bonds of the initiator have variables 
larger than $x^{sur}_c$.
In fact, if at least one descendant of $i$ had variable lower than 
$x^{sur}_c$, the avalanche would continue because this 
variable would be the minimum one on the whole perimeter. 
In order to evaluate this probability we need to take 
into account the environment of the initiator. 
In FIG.\ref{fig11} a-b we schematize 
all the possible boundary conditions for the initiator bond. 
We consider only the nearest neighbours of the initiator because,
asymptotically, the avalanches on the perimeter are influenced only by the 
environment near the zone where the avalanche evolves. 
That is, they are affected  by other branches of the aggregate which have 
some perimeter bonds affected by the avalanche. 
The presence of the boundary is implemented by allowing only the right 
and the vertical bond to grow in FIG. \ref{fig11}.

For these two cases we can evaluate the probability that the avalanche stops
immediately after the growth of the initiator, conditioned by 
the assigned boundary conditions. 
The exact value of this probability is given by the average 
of the two cases. In order to calculate the statistical weights of 
configurations (a), (b) and (c) of FIG.\ref{fig11} we use the 
void distribution $P(\lambda)$ of the random anisotropic Cantor set 
whose generators have local (near the boundary) probabilities 
$C_i;\,\,\,\,i=0,1,2$ given by the FST calculations performed in the previous 
section. 
We are then allowed to use $P(\lambda)$ with the weights obtained by FST 
because for IP the perimeter has the same statistical properties as the bulk 
of the structure. Obviously, the void distribution we obtain is a local one,  
since the probabilities $C_i$ for the Cantor set orthogonal to the 
boundary are dependent on the distance from it. 
In practice, only the $P(\lambda=0)$ can be computed with a reasonable degree
of accuracy, because it depends only upon the local properties of the set. 
When the size $\lambda$ of the void is not small with respect to the 
system size, the implicit assumption that the $C_i$ are 
independent of $z$ becomes inconsistent.

We report the expression of $P(\lambda=0)$ from \cite{FSTreport} in terms 
of $C_2$ and $C_1$:
\begin{equation}
P(\lambda=0)=\frac{C_2}{1-C_1+C_1C_2+C_1^2}.
\label{mario}
\end{equation}
The weight of configuration (a) is:
\begin{equation}
W^{(a)}=1-P(\lambda =0).
\label{Wa}
\end{equation}
The weight of configuration (b) is:
\begin{equation}
W^{(b)}= P(\lambda =0),
\label{Wb}
\end{equation}
The fixed point values of $C_2$ and $C_1$ obtained from FST 
calculation of $d^{sur}_{FST}$ in the previous section 
are $ C_2 \simeq 0.628$ and $C_1\simeq0.249$. 
If we introduce these values in eq.(\ref{mario}) we get: $P(\lambda = 0) 
\simeq 0.648$.
The probability $Q(s=1;x_c)$ to have an avalanche of duration $s=1$ is 
\[Q(s=1;x_c) = (1-x_c)^2(1-P(\lambda=0))+\]
\begin{equation}
(1-x_c)P(\lambda=0)=0.412$$
\label{mariofinale}
\end{equation} 
At this point, in order to find $\tau^{sur}$ 
we should solve the equation:
\begin{equation}
0.412 = \frac{1}{\sum_{s=1}^{\infty} s^{-\tau^{sur}}}=
\frac{1}{\zeta(\tau^{sur})}.
\label{mariozeta}
\end{equation}
\\
The numerical solution of eq.(\ref{mariozeta}) gives:
\begin{equation}
\tau^{sur}=1.55
\end{equation}
in very good agreement with our numerical findings. 
The above scheme is, however, too simplified to account for 
trapping. In fact, the method is based on the first growth step inside an 
avalanche, while trapping becomes relevant at higher orders (see Table  
\ref{tab1} and Table \ref{tab2}).

\section{Conclusions}

In this paper, we have presented, in analogy with usual critical phenomena, 
the study of boundary effects in invasion percolation 
with and without trapping. 
Near a boundary one deals with a qualitatively different rate of occupation. 
This is reflected in a lower fractal dimension of this part of the cluster.
Numerical simulations give surface fractal dimensions 
$d^{sur}=0.65 \pm 0.02$ and $d_{tr}^{sur}=0.59 \pm0.03$ 
for IP without trapping and IP with site trapping respectively. 
These two values are smaller than the bulk values. 
Meanwhile, simulations for the asymptotic shape of the 
histogram distribution and for the boundary avalanche distribution for 
IP without trapping, show that the boundary does not affect these
quantities.
The histogram self--organises into a theta function with threshold 
$x_c^{sur}=1/2$ and the boundary avalanche distribution is characterized by 
an exponent $\tau^{sur}=1.56 \pm 0.05$, very near to the bulk
value $\tau^{bulk}=1.60\pm 0.02$. 
We are also able to present a theoretical scheme to compute analytically 
the relevant critical exponents $d^{sur}$ (for both IP with and without 
trapping) and $\tau^{sur}$ (for IP without trapping only) near the boundary.
Our theoretical results $d^{sur} \simeq 0.70$, $d_{tr}^{sur}\simeq 0.66$ and 
$\tau^{sur} \simeq1.55$ are in good agreement with the numerical data. 
Authors acknowledge S. Cornell for suggestions, GC acknowledge the support of
EPSRC.

\appendix

\section{Derivation of the RTS equations}
In Invasion Percolation a bond grows at time $t$ 
if its variable is 
the minimum one at that time.
Then we can write:
$$Prob(t; (x\:\leq x_i \leq\: x+dx)\bigcap 
(x_i=\min_{k\in\partial{\cal C}_{t}}x_k)) =$$
\begin{equation}
= dx\:\:\rho_{i,t}(x)\:\prod_{k\in \partial{\cal 
C}_{t}-\{i\}}\int_{x}^{1}\:\rho_{k,t}(y)dy
\label{prob1}
\end{equation}
This gives the probability that, at time $t$, $x\:\leq x_i \leq\: x+dx$ and 
at the same time $x_i$ is the minimum on $\partial{\cal C}_{t}$ (i.e. that 
every other bond variable in $\partial{\cal C}_{t}$ is between $x$ and $1$).
By integrating eq.(\ref{prob1}) one can finally write the 
growth probability $\mu_{i,t}$ for the bond $i$ at time $t$ 
\cite{Matteo,IPlong}:
$$\mu_{i,t}\:\equiv\:Prob(t; x_i=\min_{k\in\partial{\cal 
C}_{t}}x_k) =$$
\begin{equation}
 = \int_{0}^{1}dx\rho_{i,t}(x)\:\prod_{k\in \partial{\cal 
C}_{t}-\{i\}}\int_{x}^{1}\:\rho_{k,t}(y)dy
\label{mubis}
\end{equation}
To update the effective densities $\rho_{j,t}(x)$ of generic bond not grown
$j\in \partial{\cal C}_{t}$ to obtain $\rho_{j,t+1}(x)$, we make use of the law
of conditional probability:
\begin{equation}
\label{conditio}
Prob(A|B)=\frac{Prob(A\bigcap B)}{Prob(B)}
\end{equation} 
The events $A$ and $B$ are respectively $A\equiv(x\leq x_j \leq x+dx)$ 
and $B\equiv(x_i= \min_{k\in\partial{\cal C}_{t}}x_k)$.
By definition of ``effective probability density", we can write:
\begin{equation}
Prob(t+1;  x\leq x_j\leq x+dx) = dx\:\rho_{j,t+1}(x).
\end{equation}
However, using conditional probability, we can also write
$$Prob(t+1;  x\leq x_j\leq x+dx) = $$
$$Prob(t; (x\leq x_j\leq
x+dx)\:|\:(
x_i=\min_{k\in\partial{\cal C}_{t}} x_k))=$$
\begin{equation}
\frac{Prob(t; (x\leq x_j\leq x+dx)
\bigcap( x_i=\min_{k\in\partial{\cal C}_{t}}x_k))}{Prob(t; 
x_i=\min_{k\in\partial{\cal C}_{t}} x_k)}
\label{aggiorno}
\end{equation}
The numerator of (\ref{aggiorno}) can be written as:
\[Prob(t; (x\leq x_j\leq x+dx)\bigcap( x_i=
\min_{k\in\partial{\cal 
C}_{t}} x_k)) =\]
\begin{equation}
= dx\:\rho_{j,t}(x)\int_{0}^{x}dy\:\rho_{i,t}(y)
\prod_{k\in \partial{\cal 
C}_{t}-\{i,j\}}
\left( \int_{y}^{1}du\:\rho_{k,t}(u)\right) 
\label{aggiorno1}
\end{equation}
This gives the probability that, at time $t$, $x\leq x_j\leq x+dx$,
and at the same time $x_i=\min_{k\in\partial{\cal C}_{t}}x_k)$ (i.e.
$x_i=y\in [0,x]$ and for all the other $k\in \partial{\cal C}_{t}$,
$x_k>y$.
The denominator of the right term in Eq.(\ref{aggiorno1}) is simply $\mu_{i,t}$.
Then we have:
$$\rho_{j,t+1}(x)=\frac{\rho_{j,t}(x)}{\mu_{i,t}} \int_0^x dy 
\rho_{i,t}(y) 
\prod_{k\in \partial{\cal C}_t-\{i,j\}}
\left[ \int_y^1 dz \rho_{k,t}(z) \right] 
$$

\begin{table}
\begin{centering}
\begin{tabular}{|c|c|c|c|c|} \hline
$ $ & $L=64$ &  $L=128$ & $L=192$ & $L=256$\\ \hline
$H=2$ & $0.66\pm0.02$  & $0.64\pm0.02$ & $0.63\pm0.02$ 
& $0.62\pm0.02$ \\ \hline
$H=3$ & $0.70\pm0.02$  & $0.66\pm0.02$ & $0.64\pm0.02$ & 
$0.63\pm0.02$ \\ \hline
$H=4$ & $0.73\pm0.02$  & $0.68\pm0.02$ & $0.66\pm0.02$ 
& $0.65\pm0.02$ \\ \hline
\end{tabular}
\caption{Boundary fractal dimension of IP without trapping 
for different system sizes $L$ and different values of $H$.}
\label{tab3}
\end{centering}
\end{table}

\begin{table}
\begin{centering}
\begin{tabular}{|c|c|c|c|} \hline
$ $ & $L=64$ &  $L=128$ & $L=256$\\ \hline
$H=2$ & $0.64\pm0.03$  & $0.59\pm0.03$ & $0.58\pm0.03$ \\ \hline
$H=3$ & $0.65\pm0.03$  & $0.62\pm0.03$ & $0.60\pm0.03$ \\ \hline
$H=4$ & $0.65\pm0.03$  & $0.62\pm0.03$ & $0.59\pm0.03$ \\ \hline
\end{tabular}
\caption{Boundary fractal dimension of IP with trapping 
for different system sizes $L$ and different values of $H$.}
\label{tab4}
\end{centering}
\end{table}

\begin{table}
\begin{centering}
\begin{tabular}{|c|c|c|c|c|c|c|c|c|} \hline
$n$  & $2$ &  $3$ & $4$ & $5$ & $6$ & $7$ & ...  & 
$\infty$\\ \hline
$d^{sur}(n)$  &$0.453$ & $0.576$  &$0.632$ &$0.657$  
&$0.671$  &$0.679$  &....  & $0.702$ \\ \hline
\end{tabular}
\caption{Values of the boundary fractal dimension with respect 
to the order $n$ of computation, for IP without trapping.}
\label{tab1}
\end{centering}
\end{table}

\begin{table}
\begin{centering}
\begin{tabular}{|c|c|c|c|c|c|c|c|c|} \hline
$n$ & $2$  &  $3$ & $4$ & $5$ & $6$ & $7$ & ...  & 
$\infty$\\ \hline
$d_{tr}^{sur}(n)$  & $0.453$ & $0.576$  &$0.622$ &$0.641$  
&$0.651$  &$0.657$  
&....  &$0.664$ \\ \hline
\end{tabular}
\caption{Values of the boundary fractal dimension of 
IP with site trapping with respect to the 
order $n$ of computation.}
\label{tab2}
\end{centering}
\end{table}

\clearpage
\clearpage
\end{document}